\documentclass[pre,twocolumn,superscriptaddress]{revtex4}
\usepackage[dvips]{graphicx}
\usepackage{dcolumn}
\usepackage{amssymb}
\usepackage{bm}
\begin{document}
\date{\today}
\title{Computational probes of molecular motion in the Lewis and Wahnstr\"om model for \emph{ortho}-terphenyl}
\author{Thomas G. Lombardo}
\author{Pablo G. Debenedetti}
\affiliation{Department of Chemical Engineering, Princeton
University, Princeton, NJ 08544, USA}
\author{Frank H. Stillinger}
\affiliation{Department of Chemistry, Princeton University,
Princeton, NJ 08544, USA}
\begin{abstract}
We use molecular dynamics simulations to investigate translational
and rotational diffusion in a rigid three-site model of the
fragile glass former \emph{ortho}-terphenyl, at $260 \: \text{K}
\leq T \leq 346 \: \text{K}$ and ambient pressure. An Einstein
formulation of rotational motion is presented, which supplements
the commonly-used Debye model. The latter is shown to break down
at supercooled temperatures as the mechanism of molecular
reorientation changes from small random steps to large infrequent
orientational jumps. We find that the model system exhibits
non-Gaussian behavior in translational and rotational motion,
which strengthens upon supercooling. Examination of particle
mobility reveals spatially heterogeneous dynamics in translation
and rotation, with a strong spatial correlation between
translationally and rotationally mobile particles. Application of
the Einstein formalism to the analysis of translation-rotation
decoupling results in a trend opposite to that seen in
conventional approaches based on the Debye formalism, namely an
enhancement in the effective rate of rotational motion relative to
translation upon supercooling.
\end{abstract}
\maketitle
\section{Introduction} \label{secintro}
Molecular motion in supercooled liquids has received much scrutiny
from experiments and simulations in recent years~\cite{23,12}.
These studies have uncovered a rich phenomenology not present at
temperatures above the melting point. One distinguishing feature
is commonly known as dynamic heterogeneity. This refers to the
presence of transient spatially separated regions with vastly
different relaxation times~\cite{23}. Observed in
experiments~\cite{25,38,75} and simulations~\cite{86,87,26,91,93},
these domains are separated by only a few nanometers, but can
differ by up to five orders of magnitude in their rate of
relaxation~\cite{23}. Closely related to dynamic heterogeneity is
a non-Gaussian distribution of particle displacements at times
intermediate between the ballistic and diffusive regimes of
molecular motion~\cite{26}. This behavior is a common aspect of
supercooled liquids and has been used extensively to detect
dynamic heterogeneity in computer simulations~\cite{86,26}.
Dynamic heterogeneity has also been invoked to explain the
decoupling between translational diffusion and viscosity, and
between rotational and translational diffusion in deeply
supercooled liquids~\cite{23,28,103}. At $T \gtrsim 1.2T_g$, where
$T_g$ is the glass transition temperature, the translational
diffusion coefficient, $D_t$, and the rotational diffusion
coefficient, $D_r$, are proportional to $T\eta^{-1}$, where $\eta$
is the shear viscosity. This is in accord with the Stokes-Einstein
(SE) equation for translational diffusion
\begin{equation} \label{eqse}
D_t=\frac{k_B T}{6 \pi \eta R}
\end{equation}
and its rotational counterpart, the Debye-Stokes-Einstein (DSE)
relation
\begin{equation} \label{eqdse}
D_r=\frac{k_B T}{8 \pi \eta R^3}
\end{equation}
In both expressions $k_B$ is Boltzmann's constant and $R$ is an
effective hydrodynamic radius of the diffusing particle. The fact
that these equations, which describe diffusion of a Brownian
particle in a hydrodynamic continuum with viscosity $\eta$, apply
at the molecular scale is remarkable. Yet, as liquids enter the
deeply supercooled regime, $T \lesssim 1.2T_g$, $D_t$ exhibits a
weaker temperature dependence while $D_r$ continues to adhere to
the DSE relation. Experiments on \emph{ortho}-terphenyl
(1,2-diphenylbenzene, OTP) reveal that $D_t$ is approximately
proportional to $T\eta^{-0.8}$ in this regime and that the SE
equation underpredicts $D_t$ by as much as two orders of magnitude
at $T_g + 3$~K~\cite{58,99}.

The goal of the present work is to investigate numerically the
diffusive phenomena in a model supercooled liquid. Special
emphasis is placed on rotational motion, an important aspect of
supercooled liquid dynamics that has received comparatively less
attention than its translational counterpart in existing
investigations of dynamic heterogeneity and non-Gaussian behavior.
Examples of recent studies of rotational dynamics in supercooled
liquids include refs.~\cite{127,126,124,125,88}. Inclusion of
rotational degrees of freedom provides the means to study
computationally translation-rotation decoupling, a phenomenon
observed experimentally in many fragile glass formers. To
investigate these topics we perform molecular dynamics simulations
of the rigid three-site Lewis and Wahnstr\"om model of
OTP~\cite{20} at temperatures spanning the warm thermodynamically
stable liquid to deeply supercooled states.

This paper is organized as follows. Section~\ref{secrotformalism}
provides the background on the two formalisms used here to
characterize rotational diffusion; Section~\ref{secngdh} extends
metrics for dynamic heterogeneity and non-Gaussian behavior of
translational motion to rotation. Section~\ref{secresults}
presents the results obtained from molecular dynamics simulations
of OTP and discusses their significance. The major conclusions
arrived at in this work and the open questions arising as a result
of our study are listed in Section~\ref{secdiscussion}.
\section{Rotational Formalism} \label{secrotformalism}
The most common framework for exploring rotational motion
originates with the work of Debye~\cite{95}. The underlying
physical picture views rotational diffusion as a succession of
small, random processes. A unit vector fixed to the center of mass
of a rotating molecule would then undergo a random walk on the
surface of a sphere. Solving the differential equation for the
evolution of the probability $P(\psi,t)$ that a molecule
experiences a net angular displacement $\psi$ during a time $t$
then yields~\cite{74}
\begin{equation} \label{eqdebyerotation1}
P(\psi,t) = \sum_{l=1}^{\infty}\bigg(\frac{2l+1}{2}\bigg)
P_l[\cos\psi(t)]e^{-l(l+1)D_rt}
\end{equation}
Here $P_l$ is the $l^{\text{th}}$ Legendre polynomial and $D_r$ is
the rotational diffusion coefficient, with units of inverse time.
The angular displacement is defined as
$\psi(t)=\cos^{-1}[\vec{u}(t) \cdot \vec{u}(0)]$, where $\vec{u}$
is the unit vector fixed in the molecular frame. The first two
Legendre polynomials, $l=1$ and $l=2$, can be related to several
experimental techniques including infrared absorption, Raman
scattering, and NMR~\cite{55}. These experiments often report
rotational correlation times, $\tau_l=[l(l+1)D_r]^{-1}$, in place
of a diffusion coefficient. These are straightforwardly calculated
from
\begin{equation} \label{eqdebyecorrelationtime}
\tau_l=\int_0^{\infty} \langle P_l[\cos \psi(t)] \rangle \,
\text{d}t
\end{equation}
At low temperatures, $P_l[\cos\psi(t)]$ decays slowly and long
simulations are needed to reach the time necessary to accurately
determine $\tau_l$ from equation~\ref{eqdebyecorrelationtime}. One
may derive an equivalent relation from
equation~\ref{eqdebyerotation1} and obtain a rotational
correlation time from
\begin{equation} \label{eqdebyecorrelationtime2}
\tau_l^{-1}=-\frac{\text{d}}{\text{d}t} \ln \langle
P_{l}[\cos\psi(t)] \rangle
\end{equation}
in the region where $\ln \langle P_{l}[\cos\psi(t)] \rangle$ is
linear in time. We have verified that
equations~\ref{eqdebyecorrelationtime}
and~\ref{eqdebyecorrelationtime2} give similar values of $\tau_l$
at high temperature where $P_l[\cos\psi(t)]$ decays quickly. We
refer to this formulation of rotational motion as the Debye model.

The Debye model is well-suited for examining the rotation of
dipoles and diatomic or other rigid linear molecules because these
systems provide a natural choice for the unit vector $\vec{u}$.
However, in molecules with more than two rotational degrees of
freedom (any rigid non-linear molecule), at least two orthogonal
unit vectors are required for a full description of rotational
motion. As will be shown below, rotational motion along different
directions can exhibit widely differing characteristics at low
enough temperatures. Additionally, the Debye model has limited
utility in examining dynamic heterogeneity and non-Gaussian
behavior because the angular displacement, $\psi$, is bounded
between 0 and $\pi$. As will be shown in Section~\ref{secngdh},
not only is an unbounded displacement critical to studying dynamic
heterogeneity, but rotational motion in the deeply supercooled
regime deviates appreciably, at least for the Lewis and
Wahnstr\"om model considered here, from the physical picture of
small uncorrelated angular displacements underlying the Debye
model. In light of these facts, the following alternative
rotational formalism is introduced. An angular displacement may be
obtained in a manner analogous to that of translational
displacement by integrating the angular velocity vector~\cite{54}.
\begin{equation} \label{eqrotmotion}
\Delta\vec{\varphi}_i(t) = \vec{\varphi}_i(t)-\vec{\varphi}_i(0) =
\int_0^t \vec{\omega}_i(t') \, \text{d}t'
\end{equation}
Here, $\vec{\omega}_i$ and $\Delta\vec{\varphi}_i$ are the angular
velocity and angular displacement vectors for particle $i$
respectively. Note that $\Delta\vec{\varphi}_i$ is unbounded. For
a generic rigid body with three rotational degrees of freedom each
component describes rotation about a specific principal axis fixed
in the molecular frame and originating at the center of mass. The
components are denoted by: $\vec{\omega}_i =
[\omega_i^x,\omega_i^y,\omega_i^z]$ and $\Delta\vec{\varphi}_i =
[\Delta\varphi_i^x,\Delta\varphi_i^y,\Delta\varphi_i^z]$.

\begin{figure}
\includegraphics[width=3.35in]{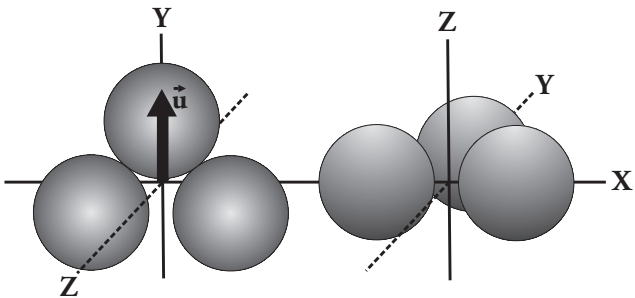}
\caption{\label{figotpmodel} Illustration of rotational directions
in the Lewis and Wahnstr\"om model of OTP. In this model each
phenyl ring is represented by a Lennard-Jones site, and the three
sites constitute a rigid isosceles triangle with a vertex angle of
$75^{\circ}$~\cite{20}. The unit vector ($\vec{u}$) used here to
study rotation according to the Debye model is shown.}
\end{figure}

In this approach, rotational motion is anisotropic; each direction
describes a specific rotational movement of the molecule.
Associated with each rotational direction is a principal moment of
inertia. These in general are not equivalent, and it is therefore
appropriate to define diffusion coefficients for each degree of
freedom, rather than for the molecule as a whole. As a starting
point, consider the Einstein relation for the translational
diffusion coefficient, $D_t$, in one dimension.
\begin{equation} \label{eqtrandiffusion}
D_t = \lim_{t \rightarrow \infty} \frac{1}{2tN} \sum_{i=1}^N
\langle [ \Delta x_i(t)]^{2} \rangle
\end{equation}
Here $\Delta x_i(t)$ is the displacement of particle $i$ in the
$x$-direction at time $t$, and the sum is over all molecules. A
rotational diffusion coefficient in the
$\varphi^{\alpha}$-direction, $D_r^{\alpha}$, can be defined
analogously as
\begin{equation} \label{eqrotdiffusion}
D_r^{\alpha} = \lim_{t \rightarrow \infty} \frac{1}{2tN}
\sum_{i=1}^N \langle [\Delta \varphi_i^{\alpha}(t)]^{2} \rangle
\end{equation}
where $\alpha \in [x,y,z]$. We refer to this description of
rotational motion as the Einstein formulation, but note that in
the rotational case the directions $\varphi^x$, $\varphi^y$, and
$\varphi^z$ are in the molecular frame. An interesting feature of
this formulation is that by appropriately selecting the
orientational axes fixed in the molecular frame, the different
diffusion coefficients will correspond to specific changes in
orientation of the molecule. The axes chosen in this paper are
illustrated in Figure~\ref{figotpmodel}.

Each of the above models of rotational motion has its advantages.
The Debye model forms the basis of several experimental techniques
(primarily via measurements of $P_2[\cos\psi(t)]$), and has
therefore played an important role in studies of
translation-rotation decoupling~\cite{117,58,97,99}. In contrast,
the Einstein formulation has only been used in simulations
\cite{54,88,111,119,126}, but as will be shown below, it offers
important advantages for the numerical investigation of dynamics
in deeply supercooled liquids, where the physical basis of the
Debye approximation appears to break down.
\section{Non-Gaussian Metrics and Dynamic Heterogeneity} \label{secngdh}
The notion of dynamic heterogeneity implies the existence of a
time scale intermediate between the ballistic and diffusive
regimes over which particles of like mobility are spatially
correlated. Therefore, in order to study dynamic heterogeneity one
must quantify ``mobility'' and identify this time scale. Most
studies have defined mobility starting from the individual
particle displacement vector $\Delta \vec{r}_i(\Delta t)$ and its
associated scalar quantity $\Delta r_i(\Delta t)$. To facilitate
comparison between translational and rotational diffusion, and to
connect with a recently developed diffusion formalism~\cite{13},
we elect to utilize the Cartesian component of the displacement
vector, $\Delta x_i(\Delta t)$ for translation (the $y$ and $z$
directions are equivalent and are included as independent samples
in all our calculations), and the individual angular displacements
$\Delta \varphi_i^{\alpha} (\Delta t)$, for rotation, as the
quantities of interest.

The work of Kob \emph{et al}.~\cite{26} revealed that the dynamic
heterogeneity of translational motion observed in computer
simulations of a binary glass-former is associated with a
non-Gaussian probability distribution of particle displacements.
For short times displacement may be approximated by $\Delta x_i
\approx v_i^x \Delta t$ resulting in a probability distribution
that is Gaussian due to the Maxwell-Boltzmann distribution of
velocities.
\begin{equation} \label{eqtransvelprobdist}
P(\Delta x_{i}) \cong \sqrt{\frac{m}{2 \pi k_B T (\Delta t)^2}}
\exp{\bigg(\frac{-m(\Delta x_i)^2}{2 k_B T (\Delta t)^2}\bigg)}
\end{equation}
Also, at sufficiently long times diffusion is purely random and
the displacement distribution asymptotically adopts a Gaussian
form.
\begin{equation} \label{eqtransprobdist}
P(\Delta x_i) \sim \frac{1}{2(\pi D_t \Delta t)^{1/2}}
\exp{\bigg(\frac{-(\Delta x_i)^2}{4 D_t \Delta t}\bigg)}
\end{equation}
At intermediate times where these approximations are invalid, the
behavior of $P(\Delta x_i)$ has been shown to be substantially
non-Gaussian, reaching a maximum deviation from the Gaussian form
at a time to be denoted by $\Delta t^*$. This time serves as the
appropriate time scale on which to study dynamic heterogeneity of
translational motion, and corresponds approximately to the
beginning of the long-time diffusive regime.

The extension of this idea to rotational motion is
straightforward. At short times angular displacement may be
approximated by $\Delta \varphi_i^{\alpha} \approx
\omega_i^{\alpha} \Delta t$ and the probability distribution of
angular displacements is accordingly Gaussian due to the
Maxwell-Boltzmann distribution of angular velocities.
\begin{equation} \label{eqrotsvelprobdist}
P(\Delta \varphi_i^{\alpha}) \cong \sqrt{\frac{I_{\alpha}}{2 \pi
k_B T (\Delta t)^2}} \exp{\bigg(\frac{-I_{\alpha}(\Delta
\varphi_i^{\alpha})^2}{2 k_B T (\Delta t)^2}\bigg)}
\end{equation}
Here $I_{\alpha}$ is the moment of inertia for a rotation in the
$\varphi^{\alpha}$-direction with $\alpha \in [x,y,z]$. In analogy
to translation, the distribution of angular displacements also
converges to a Gaussian form at long times.
\begin{equation} \label{eqrotprobdist}
P(\Delta \varphi_i^{\alpha}) \sim \frac{1}{2(\pi D_r^{\alpha}
\Delta t)^{1/2}} \exp{\bigg(\frac{-(\Delta
\varphi_i^{\alpha})^2}{4 D_r^{\alpha} \Delta t}\bigg)}
\end{equation}
Similarly to translation, we will show that at times intermediate
between these approximations $P(\Delta \varphi_i^{\alpha})$ is
non-Gaussian, and we will focus on the time of maximum
non-Gaussian behavior $\Delta t^*$. We use this time to examine
dynamic heterogeneity of rotational motion. The notation $\Delta
t^*$ is used here generically, and we will see that $\Delta t^*$
for translation is not in general the same as $\Delta t^*$ for
rotation.

Determination of $\Delta t^*$ requires an appropriate metric of
non-Gaussian behavior. This is commonly done by using the ratio of
the second and fourth moments of the appropriate probability
distribution of particle displacement~\cite{90}. A useful
non-Gaussian parameter is
\begin{equation} \label{eqtrana2}
\alpha_2^t(\Delta t)=\frac{\langle [\Delta x(\Delta t)]^{4}
\rangle}{5\langle [\Delta x(\Delta t)^{2}]
\rangle^{2}}-\frac{3}{5}
\end{equation}
for translation, and
\begin{equation} \label{eqrota2}
\alpha_2^r(\Delta t)=\frac{\langle [\Delta \varphi^{\alpha}(\Delta
t)^{4}] \rangle}{5\langle [\Delta \varphi^{\alpha}(\Delta t)^{2}]
\rangle^{2}}-\frac{3}{5}
\end{equation}
for rotation. Each parameter is defined such that Gaussian
behavior yields $\alpha_2^t=\alpha_2^r=0$. This can be easily
verified from equations~\ref{eqtransprobdist} and
\ref{eqrotprobdist}. The time at which $\alpha_2^t$ and
$\alpha_2^r$ reach their maximum value identifies $\Delta t^*$ for
translation and rotation, respectively.

Having specified a time scale, $\Delta t^*$, particles may be
classified as ``mobile'' and ``immobile'' using an appropriate
displacement cutoff. Objective specification of this cutoff is
challenging and previous studies have explored several options. We
elect to use a method developed by Shell \emph{et al}.~\cite{86}
which determines a displacement cutoff that can be easily adapted
to the rotational formalism used here. This technique fits a
two-Gaussian function of the form
\begin{equation} \label{eqtwogaussian}
P(z)=fG(z;\sigma_1) + (1-f)G(z;\sigma_2)
\end{equation}
to the probability distribution of displacements at the time
$\Delta t^*$. Here $f$ is a weighting parameter with $0 \leq f
\leq 1$, $G(z;\sigma)$ is a Gaussian distribution in $z$ with zero
mean and standard deviation $\sigma$, and $z$ signifies $\Delta x$
or $\Delta \varphi^{\alpha}$. This functional form reproduces the
measured distribution at $\Delta t^*$ well with only a slight
underestimation of the tails. A critical displacement value can
then be defined from the standard deviations of the two-Gaussian
fit. For translation, a reasonable three-dimensional cutoff is
$\Delta r^* = \sqrt{3}(\sigma_1 + \sigma_2)/2$~\cite{86}. The
$\sqrt{3}$ factor accounts for calculation of the standard
deviations from a one-dimensional probability distribution, since
$\langle [ \Delta r ]^2 \rangle = 3\langle [ \Delta x ]^2
\rangle$. For each rotational direction, a one-dimensional cutoff
is defined as $\Delta \varphi^{\alpha *}=(\sigma_1 + \sigma_2)/2$.
Particles are classified as translationally mobile if $\Delta
r_i(\Delta t^*) > \Delta r^*$ and rotationally mobile in the
$\varphi^{\alpha}$-direction if $\Delta \varphi_i^{\alpha}(\Delta
t^*)
> \Delta \varphi^{\alpha *}$.
\section{Simulation Results and Discussion} \label{secresults}
We have performed molecular dynamics simulations of the Lewis and
Wahnstr\"om model for OTP~\cite{20}. In this model each benzene
ring is represented by a single Lennard-Jones (LJ) site on a rigid
isosceles triangle. The two short sides of the triangle are one LJ
diameter, $\sigma$, in length (0.483 nm), and the long side is
1.217 LJ diameters long (0.588 nm), giving a vertex angle of
$75^{\circ}$. The sites on different molecules interact
pairwise-additively with a LJ interaction energy of
$\epsilon/k_B=600$~K. The chosen directions of the molecular frame
give $I_x:I_y:I_z=1:1.77:2.77$. Calculation runs are performed in
the \emph{NVE} ensemble for $N = 324$ molecules (972 LJ sites) at
a range of temperatures of $260 \: \text{K} \leq T \leq 346 \:
\text{K}$. The density at each temperature is given in
Table~\ref{tabdiffusion}
\begin{table}
\caption{\label{tabdiffusion} Diffusion coefficients at each
investigated temperature and density for translation and rotation
as measured via the Einstein model and the second Legendre
polynomial.}
\begin{ruledtabular}
\begin{tabular}{c c c c c c c}
$T$ & $\rho$ & $D_t$ & $D_r^x$ & $D_r^y$ & $D_r^z$ & $D_r$ ($P_2$)\\
$[\text{K}]$ & $[\text{g}/\text{cm}^3]$ & $[10^{-5}\;\text{cm}^2/\text{s}]$ & $[\text{ns}^{-1}]$ & $[\text{ns}^{-1}]$ & $[\text{ns}^{-1}]$ & $[\text{ns}^{-1}]$\\
\hline
346 & 1.027 & $0.23\quad$ & $7.6\;\:$ & $3.7\;\:$ & $3.6\;\:$ & $4.0\quad$ \\
305 & 1.055 & $0.057\;\:$ & $2.3\;\:$ & $1.3\;\:$ & $1.3\;\:$ & $1.0\quad$ \\
291 & 1.065 & $0.023\;\:$ & $0.95$ & $0.51$ & $0.58$ & $0.38\;\:$ \\
275 & 1.076 & $0.0072$ & $0.42$ & $0.29$ & $0.35$ & $0.12\;\:$ \\
266 & 1.079 & $0.0018$ & $0.18$ & $0.16$ & $0.24$ & $0.024$ \\
260 & 1.082 & $0.0013$ & $0.14$ & $0.15$ & $0.18$ & $0.014$ \\
\end{tabular}
\end{ruledtabular}
\end{table}
and corresponds to the equilibrium density for the model at 1 bar
as determined by Rinaldi \emph{et al}.~\cite{35}. With the
exception of quantities reported in Table~\ref{tabdiffusion}, all
values are reported in reduced units with the mass of an OTP
molecule, characteristic interaction energy $\epsilon$, and atomic
diameter $\sigma$ as the reference units. The reduced time unit is
then 3.19 ps. The rigid body equations of motion were integrated
using an iterative quaternion algorithm~\cite{60} with a step size
of 0.001 reduced time units. To improve statistics, multiple time
origins separated by 0.1-0.3 reduced time units were employed in
the calculations that follow.

\begin{figure*}
\includegraphics[width=7in]{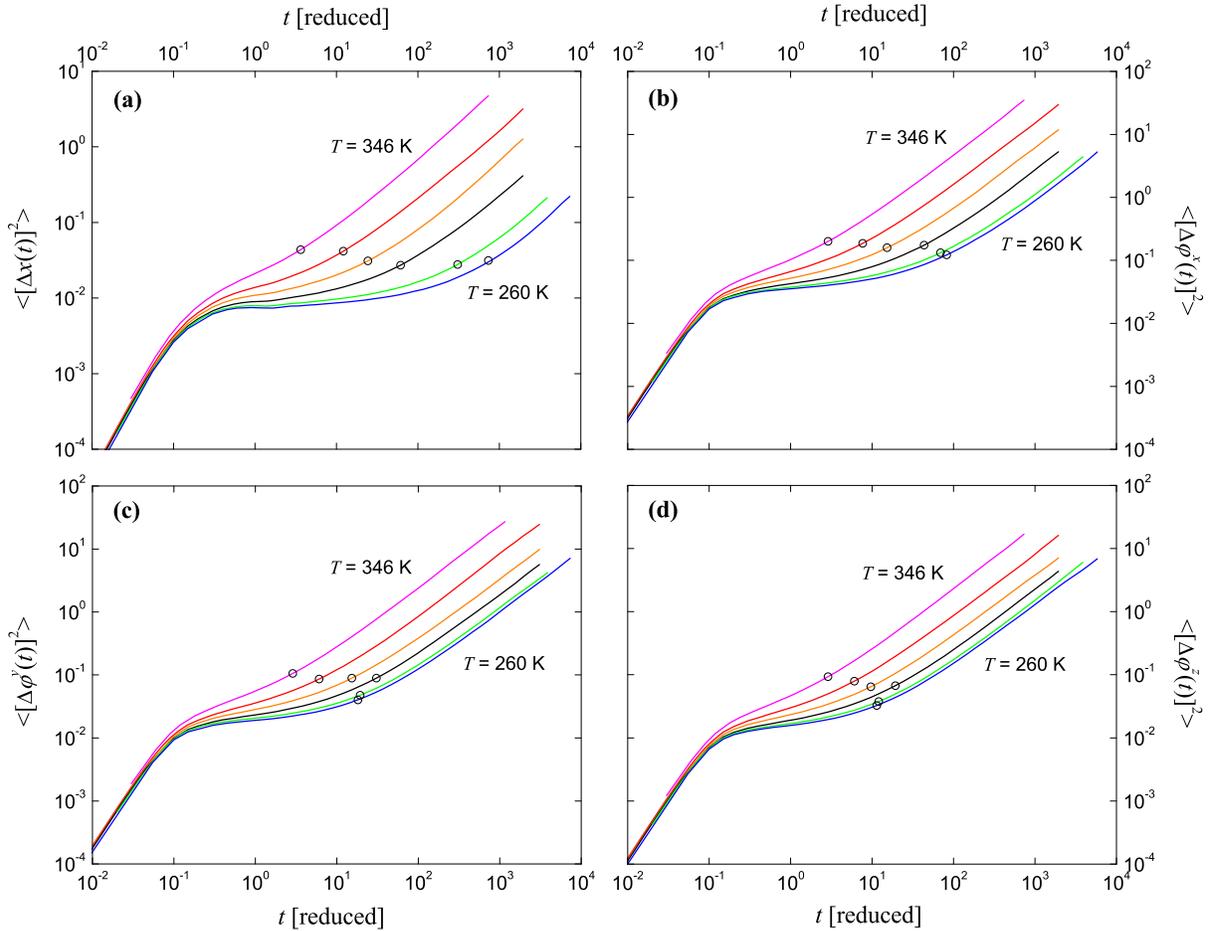}
\caption{\label{figmsd} (color online). Mean square displacement
at $T$ = 260 (blue), 266 (green), 275 (black), 291 (orange), 305
(red), and 346~K (magenta) for (a) translation in one dimension,
(b) rotation in the $\varphi^x$, (c) $\varphi^y$, and (d)
$\varphi^z$ directions. Temperature increases from bottom to top,
and the open circle on each curve marks the time of maximum
non-Gaussian behavior, $\Delta t^*$.}
\end{figure*}

The mean square displacement (MSD) is shown in Figure~\ref{figmsd}
for one-dimensional translation (the $x$, $y$, and $z$ directions
are equivalent and used as independent samples) and rotation in
each of three molecular frame directions (see
Figure~\ref{figotpmodel}). As in previous studies~\cite{54,87,88}
we observe three distinct regimes for both translational and
rotational displacement. Molecular motion is initially ballistic
with MSD proportional to $(\Delta t)^2$. Following this initial
period is a plateau corresponding to the entrapment of molecules
in the cage formed by their neighbors. Towards the end of the
plateau, the non-Gaussian parameter attains its maximum value and
thereafter the long-time diffusive regime begins, as evidenced by
the emerging proportionality between the MSD and $\Delta t$. The
diffusion coefficients for translation and rotation are determined
from the slope of the MSD in this region and are listed in
Table~\ref{tabdiffusion}. An interesting feature that emerges from
Table~\ref{tabdiffusion} is the relative value of $D_r$ between
the different rotational directions. At warm temperatures, the
rotational diffusion coefficients are ordered in accordance with
their associated moments of inertia (i.e. $D_r^x>D_r^y>D_r^z$,
consistent with $I_x < I_y < I_z$). This trend gradually reverses
itself such that at 260~K, the diffusion coefficients obey
$D_r^x<D_r^y<D_r^z$.

\begin{figure*}
\includegraphics[width=7in]{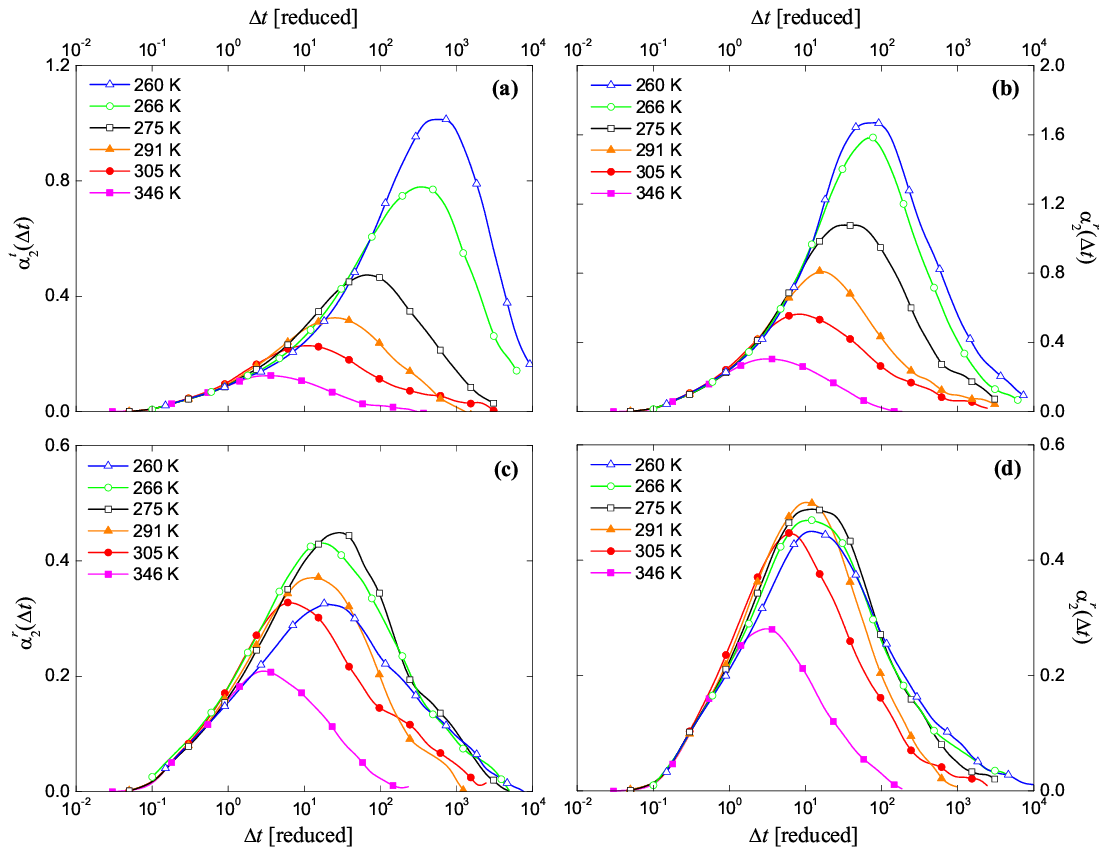}
\caption{\label{figng} (color online). The non-Gaussian parameter,
$\alpha_2$, at $T$ = 260 (blue), 266 (green), 275 (black), 291
(orange), 305 (red), and 346~K (magenta) for (a) translation in
one dimension, (b) rotation in the $\varphi^x$, (c) $\varphi^y$,
and (d) $\varphi^z$ directions. The time at which $\alpha_2$
attains its maximum value corresponds to $\Delta t^*$.}
\end{figure*}

The behavior of the non-Gaussian parameter over the range of
temperatures investigated here is shown in Figure~\ref{figng}
for translation and each of the rotational directions. As the
model OTP is cooled, $\Delta t^*$ increases and corresponds
approximately to the transition from the cage to the long-time
diffusive regime at each temperature. For $T > 291$~K, $\Delta
t^*$ is approximately coincident for translation and rotation.
However, for $T < 291$~K, $\Delta t^*(T)$ increases rapidly with
decreasing temperature for translation, such that at $T=260$~K,
$\Delta t^*$ is an order of magnitude larger for translation than
for any rotational direction. Two interesting results are shown in
Figures~\ref{figmsd} and~\ref{figng}. The first is the difference
in maximum $\alpha_2^r$ values between the $\varphi^x$ and the
$\varphi^y$ and $\varphi^z$ directions. The non-Gaussian parameter
in the $\varphi^x$ direction reaches a maximum value of
approximately 1.7 while $\alpha_2^r$ in the $\varphi^y$ and
$\varphi^z$ directions does not exceed 0.5. The second feature of
note is the non-monotonic behavior of $\Delta t^*(T)$ and
$\alpha_2^r(\Delta t^*)$ in the $\varphi^y$ and $\varphi^z$
directions. After an initial increase upon cooling down to 275~K
for $\varphi^y$ and 291~K for $\varphi^z$, $\Delta t^*(T)$ and
$\alpha_2^r(\Delta t^*)$ begin to decrease as the temperature
decreases. This trend continues down to the lowest temperature
studied. In contrast, a recent study of SPC/E water \cite{88}
showed all rotational directions to be qualitatively similar, with
$\Delta t^*(T)$ and $\alpha_2^r(\Delta t^*)$ monotonically
increasing as temperature decreases. We have verified that these
results are not an artifact of cooling at constant pressure as
opposed to constant density, by performing a subset of isochoric
runs and comparing the behavior of the non-Gaussian parameter.

A complementary approach to the study of non-Gaussian behavior is
based on a recently proposed alternative view of
self-diffusion~\cite{13}. Integration of the velocity
autocorrelation function leads to the following expressions.
\begin{equation} \label{eqavgdifftrans}
D_t=\lim_{t \rightarrow \infty} \frac{1}{3} \langle \vec{v}(0)
\cdot \Delta \vec{r}(t) \rangle
\end{equation}
\begin{equation} \label{eqavgdiffrot}
D_r^{\alpha}=\lim_{t \rightarrow \infty} \langle
\omega^{\alpha}(0) \Delta \varphi^{\alpha}(t) \rangle
\end{equation}
The physical interpretation of these equations is that the
diffusion constant is a measure of the extent to which initial
velocity biases long-time displacement.
Equations~\ref{eqavgdifftrans} and~\ref{eqavgdiffrot} can be
written more formally as an integral of a joint probability
distribution of initial velocity and final displacement~\cite{86}.
\begin{equation} \label{eqaltdifftrans}
D_t=\lim_{\Delta t \rightarrow \infty} \int \! \! \! \int v_0^x \,
\Delta x \, P(v_0^x,\Delta x) \, \text{d}v_0^x \, \text{d}\Delta x
\end{equation}
\begin{equation} \label{eqaltdiffrot}
D_r^{\alpha}=\lim_{\Delta t \rightarrow \infty} \int \! \! \! \int
\omega_0^{\alpha} \, \Delta \varphi^{\alpha} \,
P(\omega_0^{\alpha},\Delta \varphi^{\alpha}) \,
\text{d}\omega_0^{\alpha} \, \text{d}\Delta \varphi^{\alpha}
\end{equation}
Here $P(v_0^x,\Delta x)$ and $P(\omega_0^{\alpha},\Delta
\varphi^{\alpha})$ are the joint probability distributions of
initial velocity and eventual displacement, and the isotropy of
translational motion allows the replacement of $P(\vec{v}_0,\Delta
\vec{r})$ with a factor of three times the distribution of
one-dimensional displacement, $P(v_0^x,\Delta x)$. The necessary
probability distributions are easily calculated from a molecular
dynamics simulation by maintaining a two-dimensional histogram of
initial velocity and displacement after a specified time, $\Delta
t$.

\begin{figure}
\includegraphics[width=3.35in]{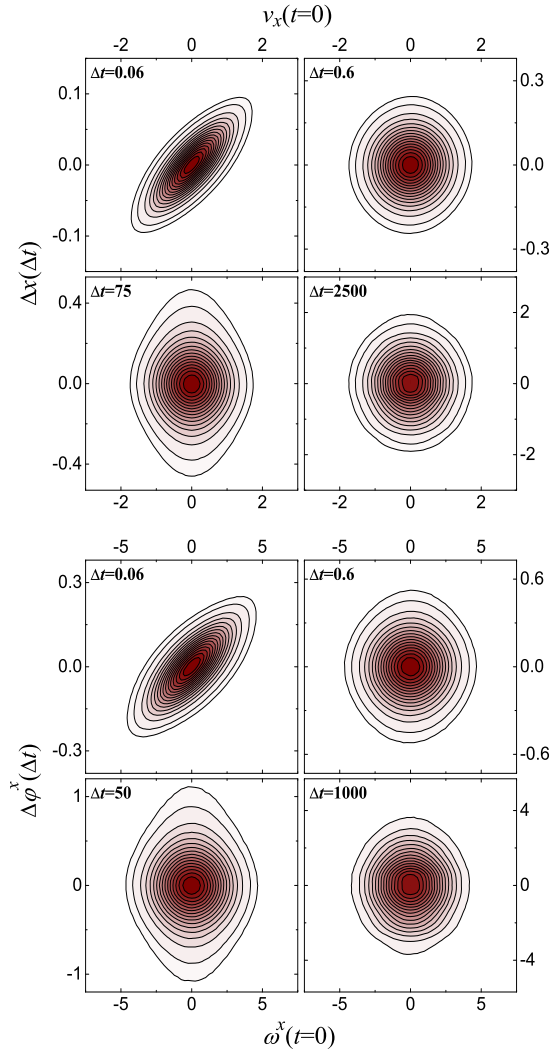}
\caption{\label{figonion} (color online). Contour plots of the
joint probability distributions of initial velocity and
displacement after a time $\Delta t$ for (top) translation and
(bottom) rotation in $\varphi^x$ at 275~K. The various values of
$\Delta t$ are marked on each plot. The $\varphi^y$ and
$\varphi^z$ directions behave similarly to the $\varphi^x$ plots
shown.}
\end{figure}

Figure~\ref{figonion}
shows contour plots of the joint probability distributions at
short, intermediate, and long times at 275~K. At short times,
initial velocity and displacement are highly correlated, resulting
in a distribution that is skewed along $\Delta x=v_0^x\Delta t$
and $\Delta \varphi^{\alpha} = \omega_0^{\alpha} \Delta t$.
Interactions with other molecules soon weaken this correlation,
and the distribution accordingly appears axisymmetric, with
circular contours clearly visible. However, when the joint
probability distribution is evaluated at $\Delta t = \Delta t^*$,
it develops a distinct ``diamond distortion''. This behavior was
first reported for translational motion by Shell \emph{et al}. for
a binary mixture of Lennard-Jones particles~\cite{85}, but has not
previously been reported for rotational motion. This distinctive
shape may be reproduced by the combination of a Maxwell-Boltzmann
distribution of velocities and a two-Gaussian distribution of
particle displacements in the form of
equation~\ref{eqtwogaussian}. Random diffusive motion eventually
returns the distributions to a single Gaussian shape at long
times. We have calculated $P(v_0^x,\Delta x)$ and
$P(\omega_0^{\alpha},\Delta \varphi^{\alpha})$ at all investigated
temperatures and for several time scales in addition to those
shown in Figure~\ref{figonion}. Over this range of conditions we
find that the time of maximum ``diamond distortion'' corresponds
approximately to $\Delta t^*$. In addition the relative strength
of the distortion is consistent with the values of $\alpha_2^t$
and $\alpha_2^r$. Namely, the contour plots become increasingly
``diamond'' shaped at $\Delta t^*$ as the temperature is reduced
with the exception of the few instances in the $\varphi^y$ and
$\varphi^z$ directions when a decrease in temperature leads to
decrease in the maximum value of $\alpha_2^r$.

The ``diamond distortion'' of the joint probability histograms
evaluated at $\Delta t^*$ reveals interesting information about
molecular motion. The circular shape of the contours at short and
long times is a result of the Gaussian nature of the particle
displacement probability distribution, since the velocities obey
the Gaussian Maxwell-Boltzmann distribution at all times. At
intermediate times, characterized by $\Delta t^*$, the tails of
the measured probability distribution of particle displacements
are significantly increased relative to a single Gaussian
distribution with zero mean and estimated standard deviation, thus
causing the ``diamond distortion''~\cite{86}. By computing
$\alpha_2^t$ and $\alpha_2^r$ for various initial velocities, we
found that non-Gaussian behavior is uniform across all initial
velocities (both translational and angular) once the regime of
ballistic particle motion has ended, after approximately 0.4
reduced time units. This is consistent with the behavior reported
in~\cite{86} for an atomic system, and confirms that the
distinctive ``diamond distortion'' is entirely due to non-Gaussian
behavior of particle displacements, and is unrelated to the
initial velocity distribution.

\begin{figure}
\includegraphics[width=3.35in]{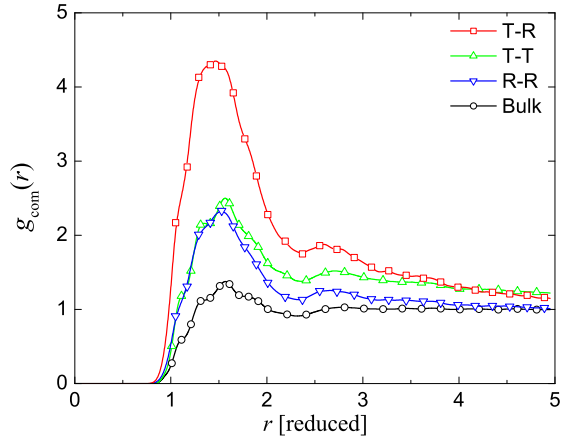}
\caption{\label{figraddist} (color online). Pair correlation
function for the centers of mass of mobile molecules as defined in
Section~\ref{secngdh} at 260~K. T-T designates the correlation
between pairs of translationally mobile molecules while R-R is the
rotational counterpart with pairs of molecules simultaneously
mobile in $\varphi^x$, $\varphi^y$, and $\varphi^z$. T-R
represents the correlation between molecules that are mobile
concurrently in translation and all rotational directions.}
\end{figure}

To detect the presence of spatially heterogeneous dynamics we have
computed the pair correlation function for the centers of mass of
mobile molecules at 260~K, as shown in Figure~\ref{figraddist}.
To compute these correlations, molecules are classified into
mobile and immobile groups for translation and each rotational
direction based on the criteria presented in
Section~\ref{secngdh}. However, the different values of $\Delta
t^*$ for translation and the three rotational directions
complicates this process. To allow the calculation of correlations
that involve more than a single direction, such as correlations
between translation and rotation or between more than a single
rotational direction, molecules are classified as mobile or
immobile based on their future displacement. This procedure begins
by selecting a particle configuration and then examining
translational and angular displacements after the appropriate
time, $\Delta t^*$, for each direction has elapsed. Molecules are
then classified as mobile or immobile in each direction, and pair
correlation functions are generated from the initial particle
configuration by examining center of mass pairwise correlations as
a function of distance between molecules that are appropriately
classified. This process is then repeated over many time origins
to improve statistics. Figure~\ref{figraddist} shows enhanced
correlations between molecules that are translationally mobile and
between molecules that are rotationally mobile. However, the
rotational correlation is only significantly increased when it is
restricted to molecules that are mobile in all rotational
directions (i.e. molecules that are simultaneously mobile in
$\varphi^x$, $\varphi^y$, and $\varphi^z$). These enhanced
correlations indicate that the dynamics of the Lewis and
Wahnstr\"om model for OTP are spatially heterogeneous in both
translation and rotation. In addition, the cross correlation
between translationally and rotationally mobile molecules (i.e.
molecules that are simultaneously mobile in translation and all
rotational directions) shows the strong tendency for these
molecules to be in close proximity. These results are similar to a
recent computational study of water, which found a strong
similarity between translational and rotational
heterogeneities~\cite{88}.

\begin{figure}
\includegraphics[width=3.35in]{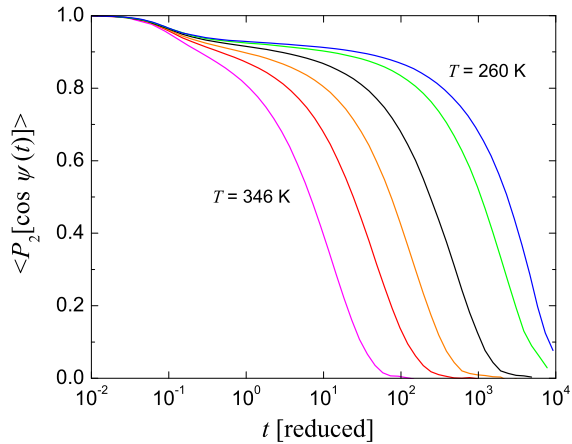}
\caption{\label{figP2} (color online). Time dependence of the
second Legendre polynomial at $T$ = 260 (blue), 266 (green), 275
(black), 291 (orange), 305 (red), and 346~K (magenta). The unit
vector, $\vec{u}$, used in this calculation is shown in
Figure~\ref{figotpmodel}. Temperature decreases moving from left
to right.}
\end{figure}

The Debye model leads to experimentally accessible measures of
rotation, and many studies of supercooled liquids, including
simulations, accordingly rely on it. Most of these examinations
obtain rotational diffusion coefficients from the decay of the
second Legendre polynomial and
equation~\ref{eqdebyecorrelationtime}. Figure~\ref{figP2}
shows the evolution of the ensemble-averaged second Legendre
polynomial calculated for the Lewis and Wahnstr\"om model of OTP
at each of the temperatures investigated. At lower temperatures a
two-step relaxation process is evident, as seen from the initial
short-time decrease, which is followed by a long-time tail. We
examined $P_1$ through $P_5$, but only show $P_2$ because of its
experimental significance. In theory, one may use any of the
Legendre polynomials to compute $D_r$ using
equation~\ref{eqdebyecorrelationtime2} and
$D_r=[l(l+1)\tau_l]^{-1}$, and should obtain the same result.
However, we find that $D_r$ decreases as the order of the Legendre
polynomial is increased (i.e $D_r(P_1)>D_r(P_2)>D_r(P_3)$ and so
on). In addition, the Debye model predicts that the ratio of
rotational correlation times measured from the first and second
Legendre polynomial, $\tau_1/\tau_2$, should be equal to 3. We
find that this ratio decreases from 2.45 at 346~K down to 1.60 at
260~K. Deviation from this theoretical value in supercooled
liquids is associated with long angular jumps~\cite{108,126}. Such
behavior was shown to be prominent in this model at 266~K by Lewis
and Wahnstr\"om~\cite{20,115}.

\begin{figure}
\includegraphics[width=3.35in]{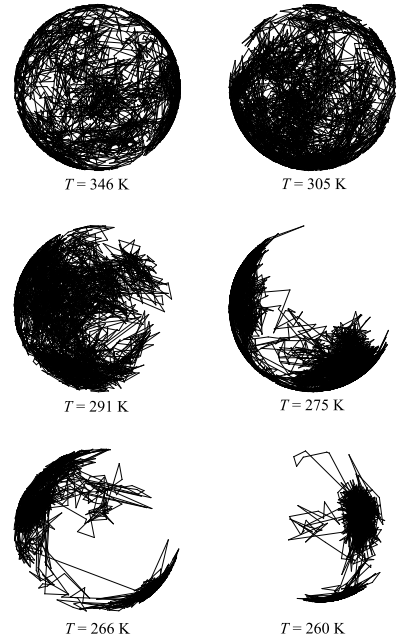}
\caption{\label{figdebyetraj} Single-molecule rotational
trajectory of the unit vector, $\vec{u}$ (see
Figure~\ref{figotpmodel}), employed in the Debye model of rotation
at each investigated temperature. The total time for each
trajectory is the time necessary for $\langle \cos\psi(t) \rangle$
to decay to a small value.}
\end{figure}

In addition to the Legendre polynomials, an informative
representation of the validity of the Debye model is the
single-molecule trajectory of the vector, $\vec{u}$, on a unit
sphere over the time required for $\langle \cos\psi(t) \rangle$ to
decay to a small value~\cite{119}. Figure~\ref{figdebyetraj}
shows representative trajectories for all investigated
temperatures. At 346~K, $\vec{u}$ explores uniformly the entire
surface of the unit sphere in a manner that is consistent with the
Debye approximation (i.e. small random steps). Upon supercooling,
the trajectory of $\vec{u}$ no longer covers the whole surface and
becomes trapped in small regions of the sphere surface over
extended periods of time. This behavior is characteristic of a
change in the mechanism of reorientation from consistently small
random steps to well-separated and sudden changes of orientation.
In this new mechanism molecules undergo librational movement for a
significant amount of time before jumping to a new orientation.
This type of movement is clearly visible in the trajectory of
$\vec{u}$ at 266 and 260~K. Figure~\ref{figdebyetraj} vividly
shows a breakdown of Debye behavior.

One of the distinctive aspects of supercooled liquid behavior is
the decoupling of translational diffusion from viscosity
(breakdown of the Stokes-Einstein equation) and of rotational
diffusion from translational diffusion~\cite{12}. It has been the
subject of many experimental (for OTP see~\cite{99,58,94,97}) and
theoretical studies~\cite{102,28,103,104}. Above the melting
temperature $D_t$ and $D_r$ are proportional to $T/\eta$. This is
in accordance with the Stokes-Einstein (SE) and
Debye-Stokes-Einstein (DSE) equations (see equations~\ref{eqse}
and~\ref{eqdse} respectively). Experiments on OTP and other deeply
supercooled fragile liquids show that the SE relation seriously
underpredicts $D_t$ whereas the DSE relation remains substantially
valid down to $T_g$~\cite{99,58,94,97}.

\begin{figure}
\includegraphics[width=3.35in]{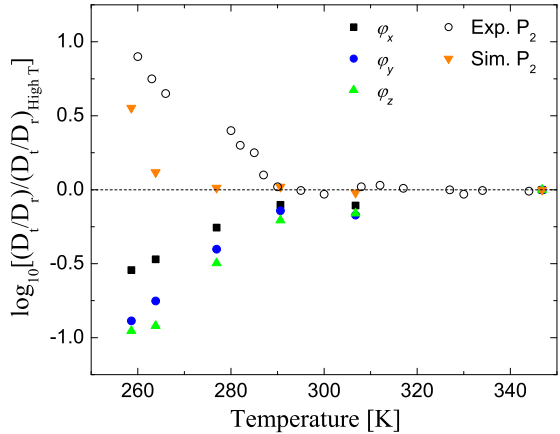}
\caption{\label{figdse} Temperature dependence of the logarithm of
the ratio of translational to rotational diffusion coefficients
normalized by the corresponding value at high temperatures.
Experimental data are from~\cite{58,99}.}
\end{figure}

A convenient metric of translation-rotation decoupling is the
quotient $D_t/D_r$ normalized by the same quantity at high
temperature~\cite{103,99} (one may equivalently use $D_t\tau_l$ in
place of $D_t/D_r$). Figure~\ref{figdse}
shows this quantity as a function of temperature, using diffusion
coefficients obtained at 346~K as the high-temperature reference
values. Included in this figure are rotational diffusion
coefficients based on the Debye model obtained from $\langle
P_2[\cos\psi(t)] \rangle$, as well as the three rotational
coefficients resulting from the Einstein formulation.
Figure~\ref{figdse} also includes experimental rotational
correlation times calculated from deuteron spin alignment
($^2\text{H-NMR}$) experiments using
equation~\ref{eqdebyecorrelationtime} with $l=2$~\cite{58,100},
and experimental translational diffusion coefficients from a
recent study by Mapes \emph{et al}.~\cite{99}.

Figure~\ref{figdse} reveals a significant result. Upon entering
the deeply supercooled regime, experiments and simulations that
measure rotation using the Debye model display a significant
increase in $D_t/D_r$ as temperature decreases. This is in accord
with the traditional concept of translation-rotation decoupling,
indicating an increase in the effective translational diffusion
coefficient relative to its rotational counterpart (i.e. the
system behaves effectively as if, upon cooling, molecules
translate further for every rotation they execute). However, when
$D_r$ is calculated from the Einstein formulation,
equation~\ref{eqrotdiffusion}, $D_t/D_r$ \emph{decreases} as the
temperature decreases, indicating an effective increase in the
rate of rotational diffusion relative to translation. This
suggests the need for a critical re-examination of our current
understanding of translation-rotation decoupling in supercooled
liquids, especially in light of its dependence on the Debye model.
\section{Concluding Remarks} \label{secdiscussion}
The diversity and complexity of dynamic phenomena present in
supercooled liquids are a major challenge to a comprehensive
understanding of this important class of condensed-phase systems.
Computer simulation is an important tool that provides insight
into the details of molecular motion in a manner that is not
currently possible in experiments. To further the understanding of
diffusion in supercooled liquids, we have studied a model of the
canonical fragile glass-former \emph{ortho}-terphenyl. In
particular, we have extended the formalism and techniques
developed for studying dynamic heterogeneity in translational
motion~\cite{86,13} to a molecular system with rotational degrees
of freedom. These methods revealed spatially heterogeneous
dynamics in translation and rotation with a strong spatial
correlation between the translationally and rotationally mobile
molecules. The commonly used Debye model of rotation was shown to
break down at deeply supercooled temperatures, as the mechanism
for molecular reorientation begins to incorporate large angular
jumps. When the Einstein formulation of rotational motion was used
to examine translation-rotation decoupling, the analysis showed a
trend opposite to that observed when using the Debye model to
quantify rotational diffusion. Specifically, the effective rate of
rotational motion appears to be enhanced relative to translation.
This result, coupled with the concurrent breakdown of the Debye
model, calls into question conventional interpretations of the
relationship between translational and rotational motion in deeply
supercooled liquids.

Models that explain translation-rotation decoupling are based on
the picture provided by dynamic heterogeneity~\cite{105,103,28}
and rely on the Debye model to describe rotation. By assuming the
presence of regions of fast and slow dynamics,
translation-rotation decoupling emerges as temperature falls below
some critical value (e.g. $T \lesssim 1.2T_g$) as a consequence of
the different ways in which translational and rotational motion
are averaged in regions of slow and fast dynamics~\cite{28}. In
this view, the SE and DSE equations are assumed to be obeyed
locally in both the slow and fast regions of the dynamically
heterogeneous liquid~\cite{25}. It can then be shown that the
effective translational diffusion coefficient is given
approximately by $(D_t^s + D_t^f)/2$, where the superscripts $s$
and $f$ denote the slow and fast regions, respectively, and it is
assumed in this approximate calculation that the slow and fast
regions each account for half of the system's volume~\cite{25}.
Using similar arguments, the rotational correlation time is given
by $(\tau_l^s + \tau_l^f)/2$. Thus, the translational diffusion
coefficient is determined by the dynamics of the fast regions,
whereas the rotational correlation time is determined by the
dynamics of the slow regions~\cite{23}. A corresponding
microscopic interpretation of the results shown in
Figure~\ref{figdse} when the Einstein formalism of rotation is
used has yet to be developed. In particular, an understanding of
how heterogeneity affects averaging in such a way as to produce an
effective enhancement of rotation upon cooling needs to be
developed.

An interesting question arising from our work is the origin of the
non-monotonic behavior of $\Delta t^*$ and $\alpha_2^r(\Delta
t^*)$ in the $\varphi^y$ and $\varphi^z$-directions [see
Figures~\ref{figng}(c) and~\ref{figng}(d)]. It is possible that
this behavior may be caused by the onset of orientational hopping
in the $\varphi^y$ and $\varphi^z$-directions. Future work will
focus on the onset of orientational hopping in the various
directions, and in particular will test the eventual appearance of
hopping in the $\varphi^x$-direction.

This and previous studies of the Lewis and Wahnstr\"om model for
OTP suggest that its behavior differs from that of real OTP at
supercooled temperatures. Its primary fault is the inaccurate
prediction of diffusion coefficients. Our study and
others~\cite{20,35} report diffusion coefficients that are three
orders of magnitude larger for translation and seven orders of
magnitude larger for rotation than experiments near 260~K
indicate. This may in part result from fitting the Lennard-Jones
interaction parameters to experimental values for the
translational diffusion coefficient and molar volume at
400~K~\cite{20} as opposed to a lower temperature. This change
could be made with relative ease, but raises the question of which
molecular features of OTP contribute most to its glass-forming
ability. OTP is known to interact with short-range van der Waals
forces~\cite{116}, and the molecular structure exhibits some
internal torsioning. These features have been incorporated into
other models for OTP, including an 18 Lennard-Jones site,
non-rigid molecule~\cite{112} and a model that uses fully
atomistic force field methods to describe the
interactions~\cite{113}. While more accurate, particularly in the
fully atomistic case, these models make it computationally
challenging to use the system sizes and simulation times necessary
at deeply supercooled temperatures. The prominence of OTP as one
of the most extensively studied fragile glass-formers attests to
the importance of developing an accurate model that captures the
salient features of real OTP but is simple enough for use in
simulation studies at supercooled temperatures. The Lewis and
Wahnstr\"om model is a first step towards this goal, but
improvements are warranted.

Our present analysis suggests further questions regarding the
nature of dynamic heterogeneity. An important open question is how
regions of high mobility emerge in the liquid. It is not yet
understood what local properties of the liquid cause the molecules
in these domains to have a high mobility. An interesting method to
explore this question was introduced by Widmer-Cooper \emph{et
al}.~\cite{118,123,122}. This technique involves running separate
simulations from a single starting configuration. At the beginning
of each run the momenta of each particle are randomly chosen from
the appropriate Maxwell-Boltzmann distribution. Their results
indicate that a particle's local environment, and not its initial
velocity, has a strong effect on its propensity for motion. This
suggests the existence of structural features that influence
particle mobility and engender dynamic heterogeneity. It would be
interesting to extend a study of this nature to include rotational
degrees of freedom. The Einstein rotational formalism used in this
paper lends itself well for such a study. Knowledge of the origin
of dynamic heterogeneity would be a significant advance in the
understanding of supercooled liquids.
\section*{ACKNOWLEDGMENTS} \label{secacknowledgments}
Useful conversations with N. Giovambattista are gratefully
acknowledged. PGD gratefully acknowledges financial support by the
US Department of Energy, Division of Chemical Sciences,
Geosciences and Biosciences, Office of Basic Energy Sciences,
Grant No. DE-FG02-87ER13714.
\bibliography{Library}
\newpage
\end{document}